\documentclass[traditabstract]{aa} % for the abstract without structuration 
\usepackage{graphicx}
\usepackage{txfonts}
\usepackage{natbib}
\bibpunct{(}{)}{;}{a}{}{,}

\newcommand{\los}{\rm{l.o.s.}}
\newcommand{\ee}{\rm{e}}

\begin{document}

\title{On the nature of IC\,3328, an early-type dwarf galaxy with weak
  spiral structure\thanks{Partly based on observations collected at
    the European Organisation for Astronomical Research in the
    Southern Hemisphere, Chile, for program 077.B-0785.}}

\titlerunning{On the nature of IC\,3328}

\author{Thorsten Lisker \and Burkhard Fuchs}
\authorrunning{Lisker \& Fuchs}

\institute{Astronomisches Rechen-Institut, Zentrum f\"ur Astronomie der
  Universit\"at Heidelberg, M\"onchhofstra\ss e 12-14, 69120
  Heidelberg, Germany\\
  \email{TL@x-astro.net}
}

%\date{Submitted to A\&A}
\date{Received date / Accepted date}

\abstract{
Various early-type dwarf galaxies with disk features have been identified in
the Virgo cluster, including
objects that display weak grand-design spiral arms despite being devoid of
gas. Are these still related to the classical dEs, or
are they a
continuation of ordinary spiral galaxies? Kinematical information of acceptable quality is
available for one of these galaxies, IC\,3328. We investigated
its dynamical configuration, taking the
effect of asymmetric drift into account, and using the Toomre parameter, as well as
 density wave considerations. The derived mass-to-light ratios and
 rotational velocities indicate a significant dynamically
 hot component in addition to the disk. However, any unambiguous conclusions
 will need to await further data for this and other
 early-type dwarfs with spiral structure.
}


\maketitle

%________________________________________________________________

\section{Introduction}
 \label{sec:intro}

As the most numerous type of galaxy in clusters and the possible
descendants of the building blocks in hierarchical structure formation,
early-type dwarf (dE) galaxies play a key role toward understanding galaxy
cluster evolution. Initially believed to be spheroidal objects having
old stellar populations and preferring the high-density regions of
clusters, today's picture of dEs is far more diverse. In the Virgo
cluster, several
subclasses with significantly different characteristics exist (Lisker
et al.\ 2007), and these are correlated with environmental
density. Those dEs that populate less dense cluster regions
partly have 
younger stellar populations \citep{p2,p4},
flatter shapes \citep{fer89,p1,p3}, and clustering
properties not like giant ellipticals but similar to spiral galaxies \citep{p3}.

\begin{figure}
\resizebox{\hsize}{!}{\rotatebox{0}{\includegraphics{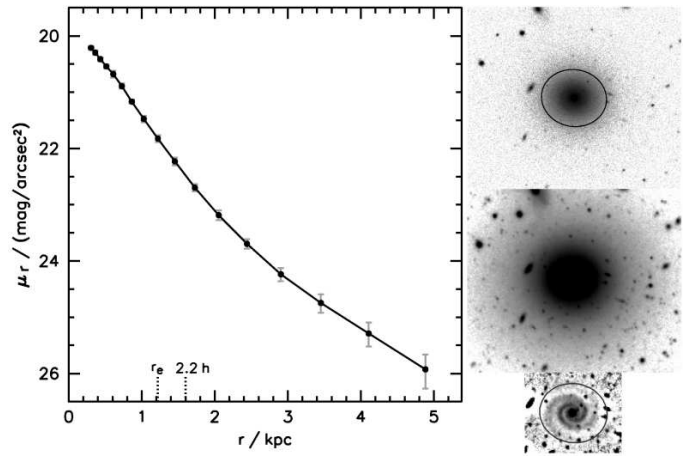}}}
\caption{{\bf Appearance and light profile of IC\,3328.} Surface
  brightness in SDSS-$r$ versus radius, adopting a distance modulus
  $m-M=31.0$\,mag \citep[$d=15.85$\,Mpc,][]{vdB96}, corresponding to a scale of
  77\,pc/$"$. The half-light radius $r_{\rm e}$, as well as 
  2.2 scalelengths $h$ (using $r_{\rm e}=1.68h$), are
 indicated by the vertical dotted lines. The images shown on the right
 are, from top to bottom, the SDSS $r$-band image, a deep white-filter
 image obtained with ESO 2.2m/WFI (program 077.B-0785), and an unsharp
 mask image of the latter, revealing the spiral arm structure. The
 half-light aperture is indicated in the SDSS and unsharp mask image.
}
\label{fig:profimage}
\end{figure}

Among them is the ``dE(di)'' subclass, characterised by
weak disk features 
-- like spiral arms or bars --
that could only be seen through unsharp masks or by subtracting
a model of the smooth galaxy light \citep{p1}. The first
discovery of spiral structure in a dE was reported by \citet{jer00a}
for the galaxy IC\,3328 (VCC\,856), which is also the focus of the work
presented here (Fig.~\ref{fig:profimage}).  The dE(di)s are not a
negligible species, but they make 
up one third of the brighter ($M_B<-16$) Virgo cluster dEs, reaching a 
fraction of 50\% at the bright end.
For those showing spiral substructure, the arm opening angles are
inconsistent with being the 
mere remainders of late-type progenitor galaxies \citep{jer00a,p1},
and the arms do not 
show up in colour maps that would indicate stellar population
differences \citep{p1}.

\begin{figure}
\resizebox{\hsize}{!}{\rotatebox{0}{\includegraphics{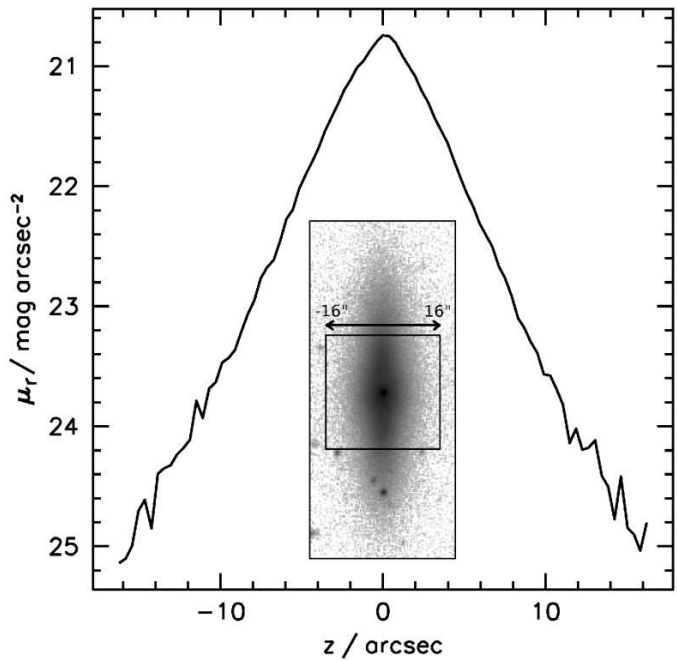}}}
\caption{{\bf Light profile in z-direction of IC\,3435.} SDSS $r$-band
  image (inset) of IC\,3435, and the intensity profile obtained by
  collapsing the outlined box-shaped region along the galaxy's major
  axis. The box has a side length corresponding to the major axis of
  the half-light elliptical isophote, or equivalently, three times its
  minor axis.  Smearing due to seeing effects can occur in the innermost
  $\sim\pm 1.5"$.
}
\label{fig:vcc1304}
\end{figure}

An indication that the dE(di)s might have been formed out of infalling disk
galaxies is that spiral arms and bar 
structures in dEs also arise in N-body simulations of galaxy
harassment \citep{mas05}, in which late-type galaxies accreted by the
cluster experience a violent structural transformation 
  (\citealt{moo96}; also see \citealt{kor09}).
 Alternatively, the dE(di)s -- being compact, gas-poor disk galaxies -- might
simply constitute the low-luminosity counterpart to normal S0/Sa
galaxies, with a few  ``dwarf-like'' S0/Sa galaxies possibly bridging
the gap to the more luminous systems (Lisker et al.\ 2006a).

%{\bf
Do the dE(di)s have a
significant dynamically hot component, which would be expected for dEs
with embedded disks, or are they consistent with
being pure disk galaxies? A first assessment can be provided by
examining an apparently edge-on dE(di), IC\,3435 (VCC\,1304;
Fig.~\ref{fig:vcc1304}; \citealt{p1}). Its axis ratio of
0.33 equals the estimate for the intrinsic axis ratio of the dE(di)s
from \citet{p3}. While it is, of course, not known
whether this galaxy hosts similar spiral structure to IC\,3328, it
still appears 
worth examining  whether a thin disk component is visible in the
vertical light profile, i.e.\ perpendicular to the disk. However,
from the SDSS $r$-band image, no such component is seen
(Fig.~\ref{fig:vcc1304}), and the
profile appears perfectly consistent with a single exponential
decline. 
%}

In
this context, it would be obvious to ask about the kinematical
properties of the dE(di)s with spiral arms. Unfortunately, useful
kinematical data is 
only available for one of these enigmatic objects, namely IC\,3328,
and only along its major axis \citep{sim02}. Here we
attempt to draw conclusions on the nature of IC\,3328
from these data.

%----------------------------------------------------------------

\section{Data}
 \label{sec:data}

\subsection{Structural measurements}
\label{sec:structure}

Total magnitude, half-light radius, and ellipticity of IC\,3328 were
measured from 
$r$-band images of the Sloan Digital Sky Survey
\citep{sdssdr5}, as outlined in \citet{p4}. The
ellipticity of 0.87, measured at the half-light radius, is consistent with the
ellipticity that could be derived from the spiral structure only, as
illustrated in Fig.~\ref{fig:profimage}. This yields an inclination
$i=30^{\circ}$. The radial surface brightness profile in $r$, shown in
Fig.~\ref{fig:profimage}, was measured from annuli of fixed elliptical
shape, using IRAF/ellipse\footnote{IRAF is distributed by the National Optical
  Astronomy Observatories, which are operated by the Association of
  Universities for Research in Astronomy, Inc., under cooperative
  agreement with the National Science Foundation.}
\citep{iraf}. Out to the radial extent of the
rotation curve (Fig.~\ref{fig:rotcurves1}) it would be compatible with an
exponential profile, i.e.\ a straight line. Nevertheless, a change in
the profile slope is seen at larger radii. This is included in the
discussion (Sect.~\ref{sec:discuss}).

\subsection{Stellar mass-to-light ratio}
\label{sec:masstolight}

From the analysis of SDSS multicolour photometry for Virgo dEs
\citep{p4}, we can derive an estimate of the stellar mass-to-light
ratio ($M/L$) of IC\,3328, using the $g-r$ and $i-z$ colours measured within
the half-light radius. With stellar population synthesis models of
\citet{bc03}, ``Padova
1994'' isochrones \citep{padova94}, a \citet{chabrier} initial mass
function, and an exponentially declining star formation rate (decay
time $\tau=1$\,Gyr), we derive luminosity-weighted age and metallicity
values of $7.5$\,Gyr (time since the onset of star formation) and
$Z=0.004$ (corresponding to $\mathrm{[Fe/H]}=-0.64$). When using
simple stellar population (SSP) models instead, the age changes to
$5.9$\,Gyr. This perfectly agrees with the spectroscopic results
from the Lick index analysis of \citet{mic08}, who find $6.0$\,Gyr and
$\mathrm{[Fe/H]}=-0.64$ using the SSP models of \citet{vaz96}. It also
agrees with a preliminary spectroscopic analysis of ESO VLT/FORS2
spectra (Paudel et al., in prep.).

 The
corresponding stellar $M/L$ in $V$ is $(M/L)_V=1.4$. Despite the good
agreement of age and metallicity from different sources, we use the
rather large age uncertainties given by \citet{mic08}
($6^{+9}_{-4}$\,Gyr) to obtain $M/L$ uncertainties
($1.4^{+1.4}_{-0.8}$). This conservative approach is also supposed to
cover any potential $M/L$ uncertainties inherent to the modelling
process.

%VCC856:
%Colours (see macro mymodels_pointsB_withvcc856):
%7.5 Gyr age, Z=0.004 in the exp_tau1 BC03 model
%-> log(age)=9.875061
%M/L(V) = 1.4023
%
%If we plot this with SSP models (Pad1994, Chabrier), it falls again
%exactly on the Z=0.004 track, and has an age of 5.9 Gyr
%-> look at log(age) = 9.778152
%M/L(V) = 1.3776
%
%This is in perfect agreement with Michielsen et al. 2008:
%6.0 +9/-4 Gyr, Fe/H = -0.64 from V96 models (SSP)
%
%Therefore, ``our'' M/L(V) value is 1.4, with a useful range
%from
% M/L(V)=0.6 (2 Gyr with BC03 SSP)
%to
% M/L(V)=2.8 (15 Gyr with BC03 SSP)

\subsection{Kinematical data}
\label{sec:kindata}

\begin{figure*}
\resizebox{\hsize}{!}{\rotatebox{0}{\includegraphics{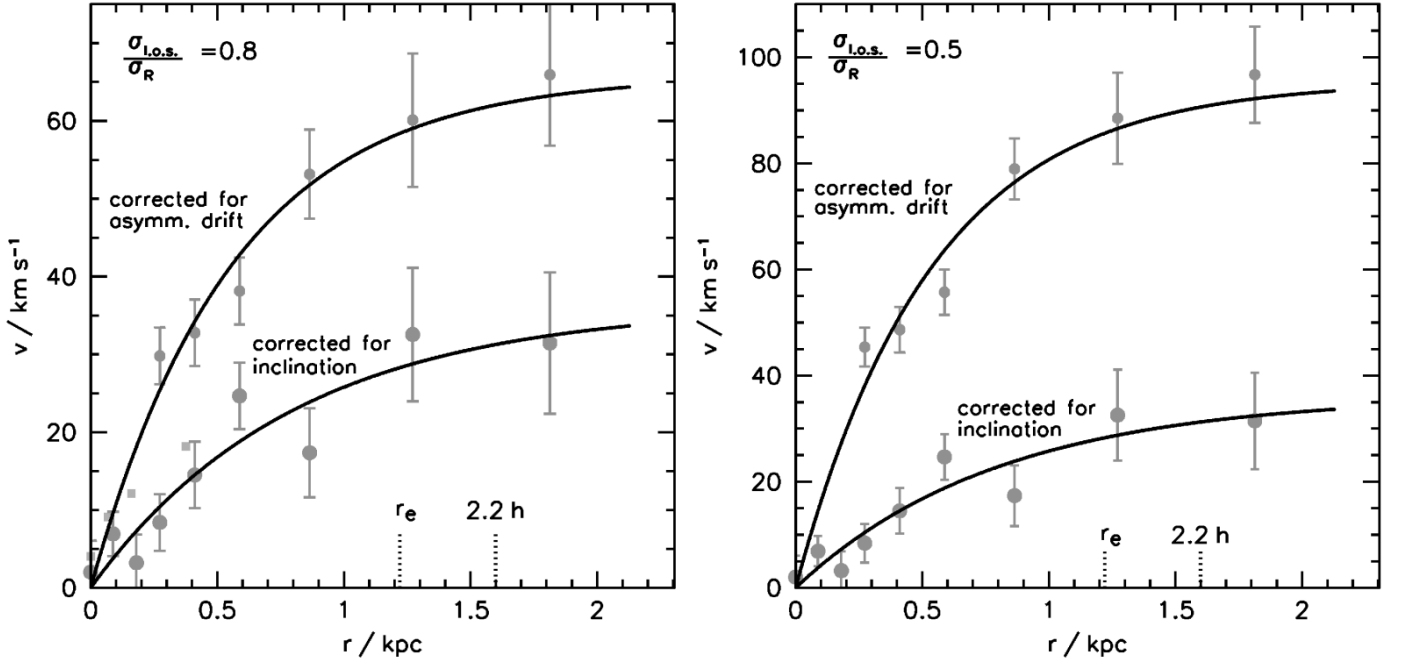}}}
\caption{{\bf Rotation curve and the effect of asymmetric drift.}
Large filled circles are the inclination-corrected measurements from
\citet{sim02}. The lower solid curve is a fit of
$f(r)=a(1-{\rm e}^{-\frac{r}{b}})$ to them. Taking the asymmetric
drift into account (see text) results in the small filled circles, along with the
corresponding fit using the same function (upper solid curve). In the left panel, we use
$\sigma_{\rm R}=45$\,km/s, while $71$\,km/s is used in the right
panel. The inclination-corrected measurements from \citet{geh03} are shown as filled squares
in the left panel.
}
\label{fig:rotcurves1}
\end{figure*}

A rotation and velocity dispersion curve of IC\,3328 was published by
\citet{sim02}, obtained from major-axis long-slit spectroscopy taken
at the Observatoire de Haute-Provence. For the analysis presented
here, the average of  published velocity data points at opposite sides
of the centre was taken, and errors propagated accordingly. We
correct for the inclination (Sect.~\ref{sec:structure})
%{\bf
and consider
the effect of finite disk thickness on the line-of-sight
velocity distribution (Appendix~\ref{sec:app_rot}),
yielding the rotation curve shown in Fig.~\ref{fig:rotcurves1}.
With the adopted intrinsic axis ratio of 0.33 \citep{p3}, the
resulting velocity correction is negligible ($<3\%$) for radii beyond 0.5 disk
scalelengths.
%}
 The rotation curve
published by \citet{geh03} is not used due to its significantly
smaller radial extent.

%----------------------------------------------------------------

\section{Kinematical analysis}
 \label{sec:kinematics}

Our approach is to analyse the kinematical data of IC\,3328 with the
assumption of a 
pure disk,
%{\bf
as implied by the vertical light profile of
IC\,3435 (Sect.~\ref{sec:intro}).
%}
We then discuss whether a
two-component galaxy would be more likely based on the results.

\subsection{Velocity dispersion}
 \label{sec:sigma}

The observed line-of-sight velocity dispersion $\sigma_{\los}$ is
more or less constant with radius, within the measurement
uncertainties \citep{sim02}, at 35\,km/s. For major axis spectra,
it can be expressed in terms of the angular and the vertical velocity
dispersion as
\begin{equation}
\label{eq:disp}
\sigma_{\los}^2 = \sigma_{\Phi}^2\sin^2i + \sigma_{\rm
  z}^2\cos^2i\quad{\rm .}
\end{equation}
Following \citet{ger00}, we adopt $\sigma_{\rm z}=0.8\sigma_{\rm R}$ as
our ``working value'', but also perform our analysis using factors of
0.4 and 0.9 instead. This range covers the values found for galaxies
of type T=1 to 5. %!!!!TBD!!!
 For $\sigma_{\Phi}$, we have
\begin{equation}
\frac{\sigma_{\Phi}}{\sigma_{\rm R}} = \frac{1}{\sqrt{2}}\sqrt{1+\frac{R}{\upsilon_{\rm c}}\frac{d\upsilon_{\rm c}}{dR}}\quad{\rm ,}
\end{equation}
which we estimate to be between 0.7 and 1. With the above range of factors, this
yields values for $\sigma_{\los}$ 
between 0.5 and 0.9 times $\sigma_{\rm R}$, leading to
$38\lesssim\sigma_{\rm R}\lesssim 71$\,km/s, with our working value
being $45$\,km/s.

\subsection{Asymmetric drift}
 \label{sec:drift}

While the asymmetric drift has a minor effect on the observed rotation
curves of normal spiral galaxies, it can be much more significant for
dwarf galaxies. The true orbital velocity at a
given radius, $\upsilon_{\rm c}(R)$, and the observed velocity, $\bar{\upsilon}_{\rm \Phi}$, are
related by the radial Jeans equation as follows \citep{binneytremaine}:

\begin{equation}
R\frac{d(\ln(\Sigma\sigma_{\rm R}^2))}{dR}+1+\frac{\upsilon_{\rm
    c}^2-\bar{\upsilon}_{\rm \Phi}^2-\sigma_{\rm \Phi}^2}{\sigma_{\rm R}^2}=0
\end{equation}
with $\upsilon_{\rm c}-\bar{\upsilon}_{\rm \Phi}$ the asymmetric
  drift velocity. (We neglect here the tilt of the velocity ellipsoid.)
Adopting $d\sigma_{\rm R}/dR=0$,
this leads to
\begin{equation}
\label{eq:asy3}
\upsilon_{\rm c}^2=\bar{\upsilon}_{\rm \Phi}^2-\sigma_{\rm
  R}^2\left(\frac{1}{2}\left(1-\frac{R}{\upsilon_{\rm c}}\frac{d\upsilon_{\rm
        c}}{dR}\right)+\frac{R}{\Sigma}\frac{d\Sigma}{dR}\right)
\end{equation}
with the surface mass density $\Sigma$.

Since we do not want to impose
any kinematical model on the analysis, we adopt the following simple method
to derive $d\upsilon_{\rm c}/dR$ and determine $\upsilon_{\rm c}(R)$. The observed,
inclination-corrected rotation curve is fitted with $f(r)=a(1-{\rm e}^{-\frac{r}{b}})$, and
its derivative is used as initial estimate for $d\upsilon_{\rm
  c}/dR$ at each data point.
The thus derived
first iteration for $\upsilon_{\rm c}(R)$ is then also fitted with
$f(r)$, and the derivative is now used as a better estimate for 
$d\upsilon_{\rm c}/dR$, applying equation~\ref{eq:asy3}.
Here, the surface mass density $\Sigma$ is taken to be the observed
luminosity density, which is obtained from the surface brightness
profile (Fig.~\ref{fig:profimage}) and the adopted distance of
$15.85$\,Mpc. No specific $M/L$ value needs to be assumed for the conversion,
since it occurs in both the enumerator and the denominator of
the last term of equation~\ref{eq:asy3}, thus cancelling out.  
In this and all following steps, we exclude data points at
$R<3"$, since the luminosity density 
is only reliable for radii significantly larger than the SDSS seeing FWHM
($\approx 1.5"$).

 This leads to the second iteration on $\upsilon_{\rm
  c}(R)$, and is repeated one more time. The resulting data
points for $\upsilon_{\rm c}(R)$ are then also fitted with $f(r)$,
yielding a rotation \emph{curve} instead of discrete data points alone. 
The
differences between the last and second-last iterations are very
small ($\lesssim 2$\,km/s).
In this procedure we always set $\upsilon_{\rm c}(0)=0$. The
resulting curves are shown in the left panel of 
Fig.~\ref{fig:rotcurves1} for our working value
$\sigma_{\rm R}=45$\,km/s (Sect.~\ref{sec:sigma}), and in the right
panel for the extreme case of $\sigma_{\rm R}=71$\,km/s. \emph{These are the
rotation curves that enter the calculations of the following sections.}

%Clearly, it would be desirable
%to obtain kinematical data of higher quality, in order to decrease the
%measurement errors, and to trace the curve to larger radii.

\subsection{Toomre stability parameter}
 \label{sec:Q}

   \begin{table}
      \caption[]{{\bf Mass-to-light ratio in $V$}, 
	using the colour transformation $V=r+0.27$
	\citep{smi02}.}
         \label{tab:ml}
     $$ 
         \begin{array}{p{0.15\linewidth}cc|c}
            \hline
            \noalign{\smallskip}
            Case      & M/L_V & M/L_V & M/L_V\,^{\mathrm{a}} \\
                      & {\scriptstyle (\sigma_{\rm R}=45\,{\rm
                          km/s})} & {\scriptstyle (\sigma_{\rm
                          R}=71\,{\rm km/s})} & {\scriptstyle (\sigma_{\rm R,Alt}=10\,{\rm km/s})} \\
            \noalign{\smallskip}
            \hline
            \noalign{\smallskip}
          % r-band:  table_r_5.txt
          %cat table_r_5.txt | awk '{printf"%5.2f  %5.2f  %5.2f\n",$1*10**(0.4*0.265),$2*10**(0.4*0.265),$3*10**(0.4*0.265)}'
            $Q=1$   & 11.2 &   25.0 & 0.9 \\ %1.0 (thin)
            $Q=1.5$ &  7.4 &   16.7 & 0.6 \\ %0.7 (thin)
            $Q=2$   &  5.6 &   12.5 & 0.4 \\ %0.5 (thin)
            $\lambda=\lambda_{\rm max}$   &   2.9 &   6.2 & 0.8 \\ % 0.8 (thin)
            \noalign{\smallskip}
	    \hline
            \noalign{\smallskip}
            Stellar & \multicolumn{3}{c}{1.4^{+1.4}_{-0.8}} \\
            \noalign{\smallskip}
            \hline
         \end{array}
     $$ 
     \begin{list}{}{}
     \item[$^{\mathrm{a}}$] The
        alternative scenario, in which only part of the observed
        velocity dispersion is attributed to the disk (see Sect.~\ref{sec:discuss}).
     \end{list}
   \end{table}
%VCC856:
%Colours (see macro mymodels_pointsB_withvcc856):
%7.5 Gyr age, Z=0.004 in the exp_tau1 BC03 model
%-> log(age)=9.875061
%M/L(V) = 1.4023
%
%If we plot this with SSP models (Pad1994, Chabrier), it falls again
%exactly on the Z=0.004 track, and has an age of 5.9 Gyr
%-> look at log(age) = 9.778152
%M/L(V) = 1.3776
%
%This is in perfect agreement with Michielsen et al. 2008:
%6.0 +9/-4 Gyr, Fe/H = -0.64 from V96 models (SSP)
%
%Therefore, ``our'' M/L(V) value is 1.4, with a useful range
%from
% M/L(V)=0.6 (2 Gyr with BC03 SSP)
%to
% M/L(V)=2.8 (15 Gyr with BC03 SSP)
%

   \begin{table}
      \caption[]{{\bf Rotational velocity}, with values at the half-light
      radius for an exponential disk and a Mestel disk, using the same
      cases as in Table~\ref{tab:ml}.}
         \label{tab:v}
     $$ 
         \begin{array}{p{0.35\linewidth}cc|c}
            \hline
            \noalign{\smallskip}
            Case      &  \upsilon_{\rm c}/km\,s^{-1}  &  \upsilon_{\rm c}/km\,s^{-1}  &  \upsilon_{\rm c}/km\,s^{-1}\,^{\mathrm{a}}  \\
                      & {\scriptstyle (\sigma_{\rm R}=45\,{\rm
                          km/s})} & {\scriptstyle (\sigma_{\rm
                          R}=71\,{\rm km/s})} & {\scriptstyle (\sigma_{\rm R,Alt}=10\,{\rm km/s})} \\
            \noalign{\smallskip}
            \hline
            \noalign{\smallskip}
            Exp., $Q=1$     &125 &188 & 38 \\ %39
            Exp., $Q=1.5$   &102 &153 & 31 \\ %32
            Exp., $Q=2$     & 89 &133 & 27 \\ %28
            Exp., $\lambda=\lambda_{\rm max}$     & 64 & 94 & 36 \\ %36
            \noalign{\smallskip}
            \hline
            \noalign{\smallskip}
            Measured + asym.\ drift & 59 & 87 & 28 \\
            \noalign{\smallskip}
            \hline
            \noalign{\smallskip}
            Mestel, $Q=1$   &113 &169 & 35 \\ %37
            Mestel, $Q=1.5$ & 92 &138 & 29 \\ %30
            Mestel, $Q=2$   & 80 &119 & 25 \\ %26
            Mestel, $\lambda=\lambda_{\rm max}$   & 58 & 84 & 34  \\ %34
            \noalign{\smallskip}
            \hline
         \end{array}
     $$ 
     \begin{list}{}{}
     \item[$^{\mathrm{a}}$] The
        alternative scenario, in which only part of the observed
        velocity dispersion is attributed to the disk (see Sect.~\ref{sec:discuss}).
     \end{list}
   \end{table}

An important diagnostic of the dynamics of a galactic disk is the
\citet{too64} stability parameter:
\begin{equation}
Q=\frac{\kappa\sigma_{\rm R}}{3.36G\Sigma}
\end{equation}
where $\Sigma$ is the surface density and
$\kappa$ the epicyclic frequency.
The value of $Q$ for a thin disk must lie in the range
$1<Q\lesssim2$, to 
ensure that the disk is not only stable against local Jeans
instabilities, but also able to develop spiral arms \citep{fuc01}.
By choosing a value for Q, the mass density $\Sigma$ can be
obtained, yielding a mass-to-light ratio when combined with the observed
luminosity density. 

%{\bf
To properly consider the disk thickness (adopted
intrinsic axis ratio 0.33, see Sects.~\ref{sec:intro} and
\ref{sec:data}), we calculate how much 
more mass a thicker disk 
contains per surface area if it has the desired properties for spiral
arm development (Appendix~\ref{sec:app_Q}). This is equivalent to using an 
``effective Q'' value that is lower than the actual one for a thin
disk. The above range of $1<Q\lesssim2$ thus changes to $0.62<Q_{\rm
  eff}\lesssim1.24$ for the case $\sigma_{\rm R}=45$\,km/s, and $0.64<Q_{\rm
  eff}\lesssim1.28$ for $\sigma_{\rm R}=71$\,km/s.

 Table~\ref{tab:ml} shows the resulting values for
$M/L$ at the half-light radius when choosing
$Q=1.0,1.5,2.0$ (i.e., $Q_{\rm eff}=0.62,0.93,1.24$ for $\sigma_{\rm
  R}=45$\,km/s, etc.).
%}
These are
compared to the stellar $M/L$ from Sect.~\ref{sec:masstolight}.

The thus derived mass density at the half-light radius can be used to
predict the rotational velocity for an exponential disk model and for a
Mestel disk. These values are listed in Table~\ref{tab:v} and illustrated
in Fig.~\ref{fig:rotcurves2}, again for 
the three $Q$ values.

\subsection{Density wave theory}
 \label{sec:lambda}

An
alternative estimate for the mass density can be obtained from density wave
theory. It predicts the circumferential wavelength $\lambda_{\rm
  max}$ at which density waves mainly grow. In terms of the
critical wavelength \citep{too64},
\begin{equation}
\label{eq:6}
\lambda_{\rm crit}=\frac{4\pi^2G\Sigma}{\kappa^2}\quad,
\end{equation}
the preferred wavelength is given by \citep{too81,ath87}
\begin{equation}
\label{eq:7}
\lambda_{\rm max}= X\left(\frac{A}{\Omega_0}\right)\cdot\lambda_{\rm
  crit}\quad .
\end{equation}
The coefficient $X$ depends on the shape of the rotation curve, measured
by Oort's constant $A$, with 
\begin{equation}
\frac{A}{\Omega_0} = \frac{1}{2}\left(1-\frac{R}{\upsilon_{\rm c}}\frac{d\upsilon_{\rm
        c}}{dR}\right)\quad ,
\end{equation}
and is given in \citet{fuc01}.\footnote{Actually, $1/X$ is given
  there.}
The number of spiral arms, $m$, is given by how often $\lambda_{\rm
  max}$ fits onto the annulus \citep{fuc08}:
\begin{equation}
\label{eq:m}
m=\frac{2\pi R}{\lambda_{\rm max}}=2\quad .
\end{equation}
From equations~\ref{eq:6}, \ref{eq:7}, and \ref{eq:m}, we have
\begin{equation}
\Sigma = \frac{\kappa^2 R}{4\pi G \cdot X\left(\frac{A}{\Omega_0}\right)}\quad .
\end{equation}

\begin{figure}
\resizebox{\hsize}{!}{\rotatebox{0}{\includegraphics{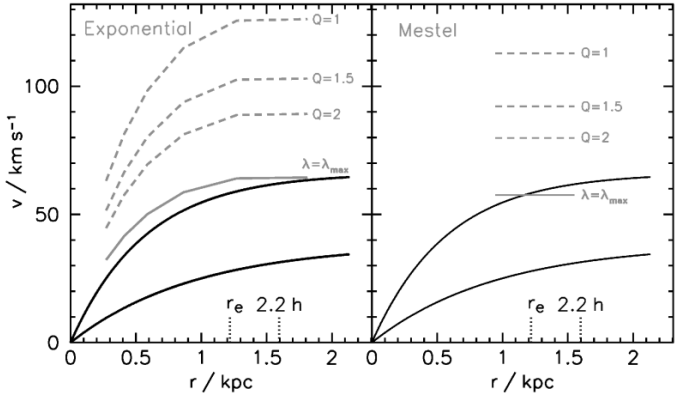}}}
\caption{{\bf Measured versus predicted rotational velocities.} The black lines are the same as in the left panel of
    Fig.~\ref{fig:rotcurves1}, i.e.\ for $\sigma_{\rm R}=45$\,km/s. The
    exponential (left) and Mestel (right) model velocities are shown
    for the cases given in Tables~\ref{tab:ml} and \ref{tab:v}, i.e.\
    $Q=1,1.5,2$ as well as $\lambda=\lambda_{\rm max}$.
}
\label{fig:rotcurves2}
\end{figure}

In contrast to the Toomre parameter, this involves no velocity
dispersion.\footnote{Since these calculations are performed based on
  the \emph{asymmetric-drift-corrected} rotation curve, a dependence
  on velocity dispersion is implicitly included through $\kappa$, since the derivation
  of $\upsilon_{\rm
    c}$ from the observed $\bar{\upsilon}_{\rm \Phi}$ involves
  $\sigma_{\rm R}$ (see equation~\ref{eq:asy3}).}
%{\bf
An approximation for taking
the actual thickness of the disk into account yields a 15\% lower value for the
coefficient $X$ (see
Appendix~\ref{sec:app_X}), which we use in the calculation of the
resulting $M/L$ values at the half-light radius (Table~\ref{tab:ml},
labelled ``$\lambda=\lambda_{\rm max}$'').
%}
Like
above for the Toomre 
parameter, rotational velocities can be predicted for an exponential and a
Mestel disk (Table~\ref{tab:v} and Fig.~\ref{fig:rotcurves2}).

%\begin{figure}
%\resizebox{\hsize}{!}{\rotatebox{0}{\includegraphics{drift_forpaper_44bc_3.eps}}}
%\caption{{\bf Estimating true rotation curves.}
%...}
%\label{fig:rotcurves3}
%\end{figure}

\section{Discussion}
 \label{sec:discuss}

%TBD kaufmann etal!

The analysis of IC\,3328 by \citet{jer00a} shows that
the spiral structure  holds clues to its dynamical
configuration. However, no rotation curve had been obtained then. With
the now available kinematical data from \citet{sim02}, a
more detailed investigation was possible.

While the mass-to-light ratios derived with the Toomre criterion
clearly exceed the stellar $M/L$ (Table~\ref{tab:ml}), the value derived
using density wave theory is still compatible with the stellar value.
The predicted rotation curve of the exponential model from $\lambda=\lambda_{\rm max}$ 
 provides
good agreement with the measured, asymmetric-drift-corrected
curve (Fig.~\ref{fig:rotcurves2}). However, as for the $M/L$, the curves derived from
the Toomre criterion lie at velocities that are too high, implying that the
picture of a pure disk is probably too simple.

For the case
of a very high velocity dispersion, the $M/L$ values are
substantially higher (columns labelled  ``$\sigma_{\rm R}=71$\,km/s'' in Tables~\ref{tab:ml}
and \ref{tab:v}), and now the value from $\lambda=\lambda_{\rm
  max}$ is also incompatible with the observed stellar range. The velocities show
the same \emph{relative} behaviour as for the lower velocity dispersion:
the model velocities from the Toomre criterion are higher than those
from $\lambda=\lambda_{\rm max}$, which agree with the measured curve.
However, the high
$M/L$ values might indicate that the actual velocity dispersion is indeed
lower than in this extreme case. Of course one could argue that a
significant amount of dark matter is present as well; and yet,
\citet{geh02} conclude that dark matter does not contribute significantly
within the half-light radius for their (admittedly small) Virgo dE sample.

The $M/L$ values from  $\lambda=\lambda_{\rm max}$ are thus the only ones that
are at least partly consistent with the inferred stellar value. A major
difference between the Toomre criterion and the density wave approach is
that velocity
dispersion does not enter the latter directly -- only implicitly
through the asymmetric drift correction -- but it does enter the Toomre
parameter directly, which yields $M/L$ values that are too high. This
might hint at a
significant dynamically hot component being 
present  in addition to the disk. If such a component contributed
substantially to the measured velocity dispersion, the disk
component's own velocity dispersion would be significantly
smaller. This would reduce the surface mass density inferred from the
Toomre criterion towards the value derived from
$\lambda=\lambda_{\rm max}$.

As for the rotational velocity,
% the observational value -- corrected for
%inclination and the asymmetric drift -- lies in between the predicted
%values for an exponential and Mestel disk when using the mass from the
%Toomre criterion on the one hand and the mass from $\lambda=\lambda_{\rm
%max}$ on the other 
%hand (Table~\ref{tab:v}). However, 
since the contribution of the
asymmetric drift is directly related to the disk's velocity
dispersion, its contribution would be much smaller in the case just
described. This would lower both the observational and model
velocities, again with particular relevance to the Toomre criterion.

%{\bf
We illustrate this possibility by repeating our analysis with a
significantly lower
velocity dispersion of $\sigma_{\rm R}=10$\,km/s
%}
and a smaller half-light
radius of 0.9\,kpc. The latter represents only the ``inner'' component of a
double-exponential fit to the light profile of Fig.~\ref{fig:profimage},
which yielded a ratio of the two components' scalelengths of 3:10, with a
central intensity ratio of 33:2. Obviously, the
asymmetric drift contribution is much smaller in this scenario. The choice of
$\sigma_{\rm R}=10$\,km/s is such that the two approaches,
$\lambda=\lambda_{\rm max}$ and the Toomre criterion, agree with each
other in the predicted rotational velocities and also the mass-to-light ratios
(Tables~\ref{tab:ml} and \ref{tab:v}, rightmost
column).\footnote{
%{\bf
Following Sect.~\ref{sec:Q} and
  Appendix~\ref{sec:app_Q}, we use here an 
``effective Q'' range of $0.7<Q\lesssim1.4$, corresponding to an
intrinsic disk axis ratio of 0.1.
%}
}
 In this case, the derived $M/L$ values
appear to be somewhat lower than the inferred stellar $M/L$ --- but,
%{\bf
given
  that now the disk contributes only part of the light that is used
  for the calculation, the values are very compatible.
%}
The model rotational velocities also agree with the
measured value. \emph{This leads to a coherent picture, in which IC\,3328
consists of a dynamically hot component with an embedded thin disk.}

Nevertheless, this interpretation must be regarded with some caution,  
given the measurement uncertainties, the moderate radial
extent of the kinematical data, and the number of assumptions and
simplifications made in the analysis.
Clearly, it would be desirable
to obtain kinematical data of higher quality, in order to decrease the
measurement errors and to trace the curve to larger radii.
Furthermore, minor axis
data would be desirable for pinning down the velocity
anisotropy.

%{\bf
Finally, we mention again that no thin disk component is seen in the 
vertical light profile of the apparently edge-on dE(di) IC\,3435 (see
Sect.~\ref{sec:intro}). It is of course not known  
whether this galaxy hosts similar spiral structure. Nevertheless, if
indeed IC\,3328 had a  
thin disk surrounded by a hotter component, then IC\,3435 would not be
an edge-on version of the same kind of object.
%}

%Even though it naturally remains unclear whether 
%
%Whether or not this weakens the above interpretation remains,
%however, unclear

\section{Summary}
 \label{sec:summary}

We attempted to gain insight into the actual nature
of early-type dwarf galaxies with weak spiral structure by analysing the
 kinematical data of IC\,3328, the prototype of this galaxy
population. To our knowledge, no other object of this kind has
kinematical data of sufficient quality.
 Based on mass-to-light ratios derived with different methods, we find
that the observed velocity dispersion of IC\,3328 cannot be
fully attributed to a stellar disk, but that a distinct dynamically hot
component is present. An unambiguous conclusion is, however, not possible,
due to the 
moderate radial coverage and the lack of minor-axis kinematical data. We
thus see our study as motivation to increase the amount of
good-quality kinematical observations for dEs.
%{\bf
%and to focus particularly on
%those dEs showing weak spiral structure, in order to understand their relation
%to ordinary dEs.
% -- diesen Nebensatz weglassen?}

%________________________________________________________________

\begin{acknowledgements}
    We thank Eva Grebel for initiating this collaboration, and the
    referee, Albert Bosma, for constructive suggestions.
    T.L.\ is supported within the framework of the Excellence Initiative
    by the German Research Foundation (DFG) through the Heidelberg
    Graduate School of Fundamental Physics (grant number GSC 129/1).
    This research has made use of the
    VizieR catalogue access tool, CDS, Strasbourg, France, and of
    NASA's Astrophysics Data System Bibliographic Services.
\end{acknowledgements}

%________________________________________________________________
 
%\bibliographystyle{aa}
%\bibliography{lisker}

%________________________________________________________________

\begin{appendix}

\section{Finite disk thickness correction of the observed rotation curve}
\label{sec:app_rot}
To obtain a first-order correction for the finite thickness
of the disk, we approximate the density profile of the galaxy with a double
exponential,
\begin{equation}
\label{eq:app1}
\nu=\ee^{-R/h_{\rm R}}\ee^{-z/h_{\rm z}} \quad ,
\end{equation}
and then integrate along the line of sight over the
density-weighted rotational velocity:
\begin{equation}
\label{eq:app2}
\upsilon_{\los}=\frac{1}{N}\int^{+\infty}_{-\infty}dl\,\nu\,
\cos(\vec{\rm e}_{\los},\vec{\rm e}_{\phi})\,
\upsilon_{\rm c} \quad ,
\end{equation}
with $N$ the normalization factor, $R=\sqrt{x^2+y^2}$, and
$l=\sqrt{y^2+z^2}$, such that the x-direction is perpendicular to the line
of sight. The unit vectors $\vec{\rm e}_{\los}$ and $\vec{\rm e}_{\phi}$
are defined in the direction of the line of sight and the direction of the
(cylindrically assumed) circular rotation of the galaxy, respectively.

We assume that the vertical scalelength $h_{\rm z}$ is much shorter
than the radial scalelength $h_{\rm R}$, and perform a Taylor expansion
of the integrand about $z=0$ to second order. After the integration, we
neglect terms higher than second order in $h_{\rm z}$, yielding
\begin{equation}
\label{eq:app3}
\upsilon_{\los}=\upsilon_{\rm c}\sin{i}
\left(1-\frac{h_{\rm z}^2}{R^2}\,tan^2i\,
\left(1-\frac{d(\ln \upsilon_{\rm c})}{d(\ln R)}\right)\right)\quad .
\end{equation}
Adopting $h_{\rm R}/h_{\rm z}=3$ \citep{p1} and using the inclination
$i=30^{\circ}$ (Sect.~\ref{sec:structure}), this results in
\begin{equation}
\label{eq:app4}
\upsilon_{\los}=\upsilon_{\rm c}\sin{i}
\left(1-\frac{1}{27}\frac{h_{\rm R}^2}{R^2}\,
\left(1-\frac{d(\ln \upsilon_{\rm c})}{d(\ln R)}\right)\right)\quad .
\end{equation}

\section{The Toomre stability criterion corrected for the finite thickness
  of galactic disks}
\label{sec:app_Q}

\begin{figure}
\resizebox{\hsize}{!}{\rotatebox{0}{\includegraphics{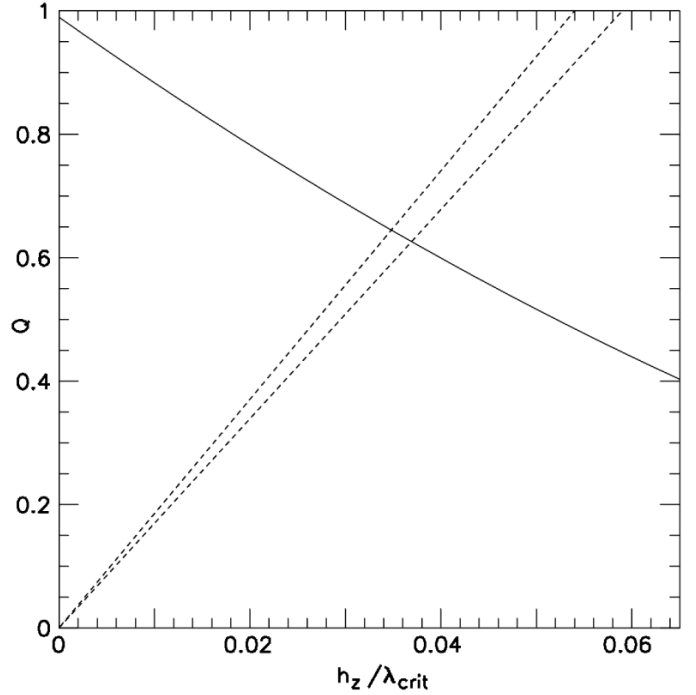}}}
\caption{The Toomre $Q$ parameter corrected for the finite thickness
    $h_{\rm z} / \lambda_{\rm crit}$ of a galactic disk. The right dashed
    line represents the case $\sigma_{\rm R}=45$\,km/s, and the left
    dashed line is for $\sigma_{\rm R}=71$\,km/s.  See the text of
    Appendix~\ref{sec:app_Q} for details.
}
\label{fig:dediqoh}
\end{figure}

The gravitational potential of a Fourier component of the surface density
of an infinitesimally thin disk is given by
\begin{equation}
\label{eq:appB1}
\Phi=-\frac{2\pi G \Sigma_{\rm k}}{|k|}\ee^{ikx-k|z|}
\end{equation}
where $k$ denotes the wave number of the Fourier term. Following
\citet{too64} we consider a disk of finite thickness as a superposition of
infinitesimally thin disks,
\begin{equation}
\label{eq:appB2}
\Phi(z)=-\frac{2\pi G \Sigma_{\rm
    k}}{|k|}\ee^{ikx}\int^{+\infty}_{-\infty}dh\,\ee^{-k|z-h|}\frac{\ee^{-|h|/h_{\rm
    z}}}{2h_{\rm z}}
\end{equation}
which leads at the midplane to
\begin{equation}
\label{eq:appB3}
\Phi(z=0)=-\frac{2\pi G \Sigma_{\rm
    k}}{|k|}\frac{1}{1+kh_{\rm z}}\quad .
\end{equation}

In the following we use the reduction term $(1+kh_{\rm z})^{-1}$, which is
slightly different from that adopted by \citet{too64}, to modify the
diagnostics of the dynamical state of a galactic disk for finite
thickness.

\citet{fuc98} show in their Appendix A1 how such a correction term must be
applied to the dispersion relation for ring-like perturbations of the
disk. In their analysis, \citet{fuc98} adopted an exponential velocity
distribution. If we switch back to a Schwarzschild distribution as assumed
by \citet{too64}, we find for the line separating exponentially growing
perturbations from neutrally stable perturbations in the space spanned by
 $\sigma_{\rm R} k_{\rm crit}/\kappa$ (or $Q$) and the
wavenumber,
\begin{equation}
\label{eq:appB4}
1=\frac{2\pi G \Sigma_{\rm k}}{\sigma_{\rm R}^2|k|}
    \left(1-\ee^{-\sigma_{\rm R}^2k^2/\kappa^2}
    I_{\rm 0}\left(\frac{\sigma_{\rm R}^2k^2}{\kappa^2}\right)\right)
    \frac{1}{1+kh_{\rm z}}
\end{equation}
where $I_{\rm 0}$ denotes the modified Bessel
function. Equation~\ref{eq:appB4} is a modified version of equation (62) of
\citet{too64}.

When $Q$ is drawn as a function of $\lambda=2\pi/k$,
we always find a lower limit for $Q$, above
which perturbations are stable on all wavelengths. This is illustrated as
a solid line in Fig.~\ref{fig:dediqoh}. As can be seen from the
figure, the allowed $Q$ values are lower
than one, which means that a disk of finite thickness can be significantly
more massive than an infinitesimally thin disk, but still be dynamically
stable.
However, it needs to be taken into account that
%, for technical reasons,
the vertical scale height $h_{\rm z}$ in units of $\lambda_{\rm crit}$
enters into the calculation. Therefore, if one increases the disk surface
density, $Q$ will drop, but $h_{\rm z} / \lambda_{\rm crit}$ drops as
well. Thus the most massive, but still dynamically stable disk model is
reached when $Q$ (originally $Q=1$) has dropped by the same fraction as
the originally calculated $h_{\rm z} / \lambda_{\rm crit}$. These models
are indicated by the intersections of the solid line and the dashed lines
in Fig.~\ref{fig:dediqoh}.

\section{Density wave theory analysis with a correction for finite thickness}
\label{sec:app_X}

We use the same reduction factor as derived in Appendix~\ref{sec:app_Q} to
correct the prediction of density wave theory for finite thickness of the
disk. Since we follow ``swing-amplification'' theory \citep{too81}, we can
straightforwardly use the formalism of \citet{fuc01}. We multiplied
the kernel $\mathcal{K}$ of his equation (68) by the reduction factor
$\left(1+h_{\rm z}\sqrt{k_{\rm x}^2+k_{\rm y}^2}\right)^{-1}$ and
again solved
the Volterra integral equation. The main result is that the
operation characteristics of the swing-amplifier mechanism, i.e.\ at which
$k_{\rm y}$ and $k_{\rm x}^{\rm eff}$ amplification peaks, are hardly affected
by this correction, the effect being of the order of 15\%. Only the
amplification amplitude is much reduced as compared to infinitesimally
thin disks. This shows
again that disks of finite thickness are dynamically more stable than
razor-thin disks.

\end{appendix}

\end{document}